\documentclass[12pt]{article}
\usepackage{cite}
\usepackage{bm}
\usepackage{dcolumn}  
\newcommand{\beq}{\begin{equation}}
\newcommand{\eeq}{\end{equation}}
\newcommand{\beqa}{\begin{eqnarray}}
\newcommand{\eeqa}{\end{eqnarray}}
\newcommand{\beqar}{\begin{eqnarray*}}
\newcommand{\eeqar}{\end{eqnarray*}}

\newcommand{\eps}{\epsilon}

\newcommand{\z}{\zeta}

\newcommand{\ie}{{\it i.e.,}\ }
\newcommand{\labell}[1]{\label{#1}} 
\newcommand{\reef}[1]{(\ref{#1})}
\newcommand\prt{\partial}

\newcommand\vtheta{\vartheta}

\begin{document}
\baselineskip 18pt%
\begin{titlepage}
\hfill
\vbox{
    \halign{#\hfil         \cr
           } 
      }  
\vspace*{20mm}
\vspace*{1mm}%
\vspace*{15mm}%

\centerline{{\Large {\bf The RN/CFT Correspondence }}}
\centerline{{\Large{\bf    }}}
\vspace*{5mm}
\begin{center}
{Ahmad Ghodsi  and  Mohammad R. Garousi}\\
\vspace*{0.2cm}
{ Department of Physics, Ferdowsi University of Mashhad, \\
P.O. Box 1436, Mashhad, Iran}\\
\vspace*{0.1cm}
\vspace*{1.1cm}
{E-mails: ahmad@mail.ipm.ir  and  garousi@mail.ipm.ir}
\end{center}

\begin{abstract} 
Recently it has been shown in 0901.0931 [hep-th]  that  the approach to extremality for the non-extremal Reissner-Nordstrom black hole is not continuous. The non-extremal RN black hole splits into two spacetimes at the extremality: an extremal black hole and a disconnected $AdS_2\times S^2$ space  which has been called the ``compactification solution". As a possible resolution for understanding the entropy of extremal RN black hole, it has been speculated  that the entropy of the non-extremal black hole may be carried by the latter solution. 
By uplifting the four dimensional ``compactification solution" with electric charge $Q_e$  to a five dimensional solution, we  show that this solution  is dual to a  CFT with central charge $c=6Q_e^3$. The  Cardy formula then shows that  the microscopic entropy of the CFT is the same as the macroscopic  entropy of the ``compactification solution". 
\end{abstract} 

\end{titlepage}


\section{Introduction} \label{intro} 
One of the most exciting  observation in the modern theoretical physics is the holographic dualities that relates a quantum gravity to a quantum field theory without gravity in fewer dimensions \cite{'tHooft:1993gx,Susskind:1994vu}. The best understood holographic duality is the duality between the ten dimensional type IIB string theory  on background $AdS_5\times S^5$ with flux and the four dimensional  ${\mathcal N}=4$ super  Yang-Mills theory at the boundary of $AdS_5$ \cite{Maldacena:1997re,Gubser:1998bc,Witten:1998qj}. 
Recently the idea of the holographic duality has been examined for the more interesting backgrounds using the Brown and Henneaux technique  \cite{Brown:1986nw}.  It has been shown in \cite{Guica:2008mu} that there is a two-dimensional CFT dual of quantum gravity  on  extreme Kerr  background. Even though the structure of the CFT is not known, the central charge of the CFT can be found by studying the nontrivial asymptotic symmetry of the extreme Kerr solution. The Cardy formula then gives the microscopic entropy of the CFT to be exactly the same as the macroscopic entropy of the extreme Kerr background \cite{Guica:2008mu}. This duality has been extended to other backgrounds in \cite{Hartman:2008pb,recent,Compere:2009dp}, (see also \cite{Solodukhin:1998tc}).

In this paper we would like to study the  holographic duality for the extreme limit of the Reissner-Nordstrom solution. It has been argued in \cite{Hawking:1994ii} that the semiclassical method gives zero result for the entropy of any extremal black hole   even if its  horizon area  is non-zero.  The reason is that the space outside  the horizon  of a non-extremal black hole is a manifold with topology $R^2\times S^2$, whereas,  the space outside the horizon of an extremal black hole is a manifold with  topology $R\times S^1\times S^2$, \ie the horizon  is excluded because  the physical distance between an arbitrary point and the horizon  is infinite. 
It has been shown in \cite{Carroll:2009ma} that the approach to extremality for RN black holes is not continuous. The non-extremal RN black hole splits into two spacetimes at the extremality: an extremal black hole and a disconnected $AdS$ space which has been called  the ``compactification solution". It has been speculated in \cite{Carroll:2009ma} that the entropy of the non-extremal RN  black hole may be carried by the ``compactification solution" when one takes  the extremal limit. 
  
In this paper we would like to  find the CFT dual of the ``compactification solution" by applying the Brown-Henneaux  technique. It has been argued in \cite{Hartman:2008pb} that  the  gauge symmetry of the extreme Kerr-Newman-AdS black hole may be  combined  with the geometry of the four dimensional extreme Kerr-Newman-AdS black hole to write a five dimensional metric from which   the central charge of the extreme RN  black hole can be found in the limit $J\rightarrow 0$. Using this  idea we  find a five dimensional solution which reduces to the four dimensional ``compactification solution" upon compactifying the 5-th dimension. The CFT dual of this five dimensional solution should be also dual to the  four dimensional solution.  
  
The paper is organized as follows. In the next section we review the non-extremal RN solution of Einstein-Maxwell theory in four dimensions and its  extremal limits. In section 3 we uplift the compactification solution to a five dimensional Einstein-Maxwell theory. In section 4, we study the CFT dual of the five dimensional solution  and show that a part of the  $U(1)$ isometry of the compactification solution appears at the boundary as Virasoro algebra with  a central charge which gives exactly the microscopic entropy after using the Cardy formula.

\section{Review of  non-extremal  RN solution}
In this section we review the non-extremal Reissner-Nordstrom solution
of the Einstein-Maxwell  theory in four dimensions. The action  is given by
\beqa 
S=\frac{1}{16\pi G}\int d^{4}x\,\sqrt{-g}\bigg\{
R
-\frac{1}{4}{\cal F}_{(2)}^2\bigg\}\,,\labell{tree}
\eeqa 
where $G$ is the four dimensional Newton's constant.

The non-extremal Reissner-Nordstrom solution with mass $M$ and  electric charge $Q_e$ is given by
\beqa
ds^2&=&-\left(1-\frac{r_+}{r}\right)\left(1-\frac{r_-}{r}\right)dt^2+\frac{1}{\left(1-\frac{r_+}{r}\right)\left(1-\frac{r_-}{r}\right)}dr^2+r^2d\Omega_2^2\,,\nonumber\\ {\cal F}_{(2)}&=&\frac{2\sqrt{G}Q_e}{r^2}dt\wedge dr\,.\nonumber
\eeqa
There are two event horizons located at the coordinate singularities
\beqa
r_{\pm}&=&GM\pm\sqrt{G^2M^2-GQ_e^2}\,.
\eeqa
 There are different types of patches
\beqa
{\rm Region I}:&&r_+<r<\infty\,,\qquad -\infty<t<\infty\,,\nonumber\\ 
{\rm Region II}:&&r_-<r<r_+\,,\qquad -\infty<t<\infty\,,\nonumber\\ 
{\rm Region III}:&&\,0\,<\,r<r_-\,,\qquad -\infty<t<\infty \,.
\eeqa
The distance between an arbitrary point and the outer horizon is finite, hence, the entropy of this  solution can be found from the semi-classical method, \ie
\beq
S=\frac{\pi r_+^2}{G}.\labell{entropy1}
\eeq 
The Hawking temperature of the black hole which is given by $2\pi T=\sqrt{g^{rr}}\prt_r\sqrt{g_{tt}}$ at the outer horizon is
\beqa
T&=&\frac{1}{2\pi r_+^2}(r_+-r_-)\,.\labell{T}
\eeqa
The Hawking temperature is zero when $r_+=r_-$, however, the entropy remains non-zero.

\subsection{Extremal limits}

It has been shown in \cite{Carroll:2009ma} that the approach to extremality for RN black holes is not continuous. The non-extremal RN black hole splits into two spacetimes at the extremality: an extremal black hole and a  disconnected  ``compactification solution".
 The  extremal black hole with event horizon at $r=Q_e$ is
\beqa
ds^2&=&-(1-\frac{Q_e}{r})^2dt^2+\frac{1}{(1-\frac{Q_e}{r})^2}dr^2+r^2d\Omega_2^2\,,\nonumber\\ 
{\cal F}_{(2)}&=&\frac{2Q_e}{r^2}dt\wedge dr,\labell{extremal}
\eeqa 
where we have set $G=1$.  There are two regions I, III for this solution. The region $II$ disappears in this limit. The physical distance between an arbitrary point and the horizon is infinite\footnote{If one consider the Reissner-Nordstrom solution as  a solution of the effective theory of the string theory, the situation will change. In that case, it has been argued in \cite{Horowitz:1996qd} that near the horizon, the length of periodic time coordinate approaches to zero and hence the string winding modes become massless or even tachyonic. So one must include these modes to the effective action. It has been speculated in \cite{Horowitz:1996qd} that in the presence of these modes the physical distance between an arbitrary point and the horizon remains finite, hence,  the macroscopic entropy of extremal solution of the string theory effective action is non-zero which should be the same as the microscopic entropy of string microstate counting \cite{Strominger:1996sh}.} so the semiclassical method gives no entropy for this solution \cite{Hawking:1994ii}. To go to the near  horizon in the region $I$, one introduces the new spacelike coordinate $0<\lambda<\infty$ and timelike coordinate $-\infty<\sigma<\infty$  as
\beqa
\lambda&=&\frac{r-Q_e}{\eps}\,,\quad \sigma\,=\,-\frac{t\eps}{Q_e^2}\,,
\eeqa
and takes the limit $\eps\rightarrow 0$ keeping $(\lambda, \,\sigma)$ fixed. The solution for arbitrary $\lambda$ becomes
 \beqa
 ds^2&=&Q_e^2\left(-\lambda^2d\sigma^2+\frac{1}{\lambda^2}d\lambda^2+d\Omega_2^2\right)\,,\nonumber\\
 {\cal F}_{(2)}&=&-2Q_ed\sigma\wedge d\lambda\,,\labell{nearH}
 \eeqa
which is  $AdS_2\times S^2$. Similar  geometry for extreme Kerr solution has been found in \cite{Bardeen:1999px}.

The ``compactification solution" on the other hand  has the three regions I, II, and III. In fact the physical distance between the inner and outer horizons of the non-extremal solution remains non-zero in this case \cite{Carroll:2009ma}. By appropriate coordinate transformation, the metric of the three regions  can be mapped to a global  $AdS_2\times S^2$  solution \cite{Carroll:2009ma}. For instance, in region II using the new timelike coordinate $0<\chi<\pi$ and spacelike coordinate $-\infty<\psi<\infty$ via  the following coordinate transformation:
\beq
r=Q_e-\epsilon\cos\chi\,,\quad\psi=\frac{\epsilon}{Q_e^2}t\,,
\eeq 
where $\epsilon=\sqrt{M^2-Q_e^2}$, one finds the metric and the field strength map to
\beqa
ds^2&=&Q_e^2(-d\chi^2+\sin^2\chi d\psi^2+d\Omega_2^2)\,,\nonumber\\
{\cal F}_{(2)}&=&2Q_e\sin\chi d\psi\wedge d\chi\,,\labell{extremal2}
\eeqa
where we have sent $\epsilon\rightarrow 0$. Note that in this limit $r_+=r_-=Q_e$ and at the same time $r\rightarrow Q_e$. Moreover, the physical distance between the outer and the inner horizons remains non zero at this limit. Using the coordinate transformation
\beqa
\cos\chi=\frac{\cos\tau}{\cos\vtheta}\,,\quad \tanh\psi=\frac{\sin\vtheta}{\sin\tau}\,,
\eeqa
the metric \reef{extremal2} transforms to \cite{Carroll:2009ma}
\beqa
ds^2&=&\frac{Q_e^2}{\cos^2\vtheta}(-d\tau^2+d\vtheta^2)+Q_e^2d\Omega_2^2\,,\labell{extremal3}
\eeqa
which is $AdS_2\times S^2$.  The flux is mapped to
\beqa
 {\cal F}_{(2)}&=&-\frac{2Q_e}{\cos^2\vtheta}d\tau\wedge d\vtheta.\labell{global}
\eeqa 
The metric \reef{extremal3} covers a portion of the global $AdS_2$. The other portions of the entire manifold are covered by the metric in regions I and III \cite{Carroll:2009ma}. The boundaries of the global $AdS_2$ are at $\vtheta=\pm\pi/2$. In terms of new coordinate $u=1/\cos\vtheta$, the boundaries are at $u\rightarrow\infty$. The solution in terms of the $u$-coordinate is
\beqa
ds^2&=&Q_e^2\left(-u^2d\tau^2+\frac{du^2}{u^2-1}+d\Omega_2^2\right)\,,\nonumber\\
{\cal F}_{(2)}&=&-\frac{2Q_eu}{\sqrt{u^2-1}}d\tau\wedge du\,.\labell{comsol}
\eeqa
Near the boundary, $u\rightarrow\infty$,  it behaves as 
\beqa
ds^2&=&Q_e^2\left(-u^2d\tau^2+\frac{du^2}{u^2}+d\Omega_2^2\right)\,,\nonumber\\
{\cal F}_{(2)}&=&-2Q_ed\tau\wedge du\,,
\eeqa
which is similar to the near horizon geometry of extremal black hole \reef{nearH}. 

It has been shown in \cite{Carroll:2009ma} that in the extremal limit, region II and the near horizon in regions I and III of the non-extremal RN black hole become the compactification solution \reef{comsol}, while the portions of regions I and III with any finite distance away from the horizon form the extremal RN black hole \reef{extremal}.
 
In the extremal limit the entropy \reef{entropy1}  remains non-zero, \ie 
\beqa
S_{\rm macro}&=&\pi Q_e^2\,,\labell{entropy2}
\eeqa
and the  Hawking temperature \reef{T} is zero, however, there is another  temperature which is conjugate to the electric charge and is defined by  $T_edS=dQ_e$. This temperature   is
\beqa
T_e&=&\frac{1}{2\pi Q_e}\,.\labell{Te}
\eeqa
The macroscopic entropy \reef{entropy2} should be extracted also  from  microstates counting. If the extremal RN black hole carries the macroscopic entropy, then the microstates counting of the CFT dual at the boundary of \reef{nearH} should give the macroscopic entropy. On the other hand, if the compactification solution \reef{comsol} carries the entropy, then the microstates counting of the CFT dual at the boundary of \reef{comsol} should give the macroscopic entropy. It has been suggested in \cite{Carroll:2009ma} that a possible resolution, for having no entropy for the  extremal RN black hole  in the semiclassical method \cite{Hawking:1994ii}, is that the macroscopic entropy \reef{entropy2}  is carried  by the compactification solution. In this paper we would like to study the CFT dual of this solution. 

\section{Uplifting to five dimensions  }

To study the 2-dimensional CFT dual of the compactification solution using the Brown-Henneaux technique \cite{Brown:1986nw} that has been used for the extreme Kerr solution in \cite{Guica:2008mu}, one should write the metric in a canonical form which has isometry $SL(2,R)\times U(1)$ with off-diagonal metric in the $U(1)$ part. Using this idea, the $U(1)$ gauge symmetry of the extreme Kerr-Newman-AdS black hole has been combined in \cite{Hartman:2008pb} with the geometry of the four dimensional extreme Kerr-Newman-AdS black hole to write a canonical five dimensional metric which has off-diagonal component in the 5th-direction. We note that the new metric  must satisfy the equations of motion in order to use the formula for the 5-dimensional on-shell generators \cite{Barnich:2001jy}.  We will show that in the present case only a part of the  the $U(1)$ gauge field \reef{global} can be combined with the metric \reef{extremal3} to write a canonical five dimensional metric which  satisfies the  equations of motion. In this section,    we  uplift the compactification solution to  the  five dimensions,   and then in section 4 find the CFT dual of the five dimensional solution. 

Consider the following five   dimensional   theory: 
\beqa 
S=\frac{1}{16\pi G^{(5)}}\int d^{5}x\,\sqrt{-g}\bigg\{
R
-\frac{1}{12}F_{(3)}^2\bigg\}\,.\labell{tree2}
\eeqa The equations of motion are
\beqa
R_{\mu\nu}&=&
\frac{1}{12}\left(3F_{\mu}{}^{\alpha\beta}F_{\nu\alpha\beta}-\frac{2}{3}g_{\mu\nu}F^2_{(3)}\right)\,,\nonumber\\
\prt_{\mu}\left(\sqrt{-g}F^{\mu\alpha\beta}\right)&=&0\,.
\eeqa
The above equations  are satisfied by the following solution:
\beqa
 ds_5^2&=&\frac{Q_e^2}{\cos^2\vtheta}(-d\tau^2+d\vtheta^2)+Q_e^2d\Omega_2^2+(dy+Q_e\tan\vtheta d\tau)^2\,,\nonumber\\
 F_{(3)}&=&\frac{\sqrt{3}Q_e}{\cos^2\vtheta}d\tau\wedge d\vtheta\wedge dy\,,\labell{sol5}
 \eeqa
where $y$ is a fibered coordinate with period $2\pi$. In the above solution, $Q_e$ is a constant which we will take  to be the four dimensional electric charge.

Upon dimensionally reducing   the $y$ coordinate as \cite{Stelle:1998xg}
\beq
ds_{d+1}^2=e^{2\alpha\phi} ds_d^2+e^{2\beta\phi}(dy+{\cal A})^2\,,
\eeq
where $\beta=(2-d)\alpha$ and $\alpha^2=1/[2(d-1)(d-2)]$, the action \reef{tree2} reduces to 
\beqa
S&=&\frac{1}{16\pi }\int d^{4}x\,\sqrt{-g}\left(R-\frac{1}{2}(\prt\phi)^2-\frac{1}{4}e^{-2(d-1)\alpha\phi}{\cal F}_{(2)}^2-\frac{1}{4}e^{2(d-3)\alpha\phi}F_{(2)}^2\right)\,,\labell{act}
\eeqa
where ${\cal F}_{(2)}=d{\cal A}$, and we have used the fact that the field strength $F_{(3)}$ has  component along the $y$-direction, \ie $F_{(2)}=dA$ is the reduction of $F_{(3)}$. The equation of motion of the dilaton is
\beqa
D^2\phi&=&-\frac{2(d-1)\alpha}{4}e^{-2(d-1)\alpha\phi}{\cal F}_{(2)}^2+\frac{2(d-3)\alpha}{4}e^{2(d-3)\alpha\phi}F_{(2)}^2\,.
\eeqa
For the present case that $d=4$, one finds  $\phi=0$ is a solution of the above equation. 

Using the $\phi=0$, the  action \reef{act} reduces to 
\beqa
S&=&\frac{1}{16\pi }\int d^{4}x\,\sqrt{-g}\left(R-\frac{1}{4}{\cal F}_{(2)}^2-\frac{1}{4}F_{(2)}^2\right)\,,\labell{act2}
\eeqa
 and the five dimensional solution \reef{sol5} reduces to the following   solution:
\beqa
ds_4^2&=& \frac{Q_e^2}{\cos^2\vtheta}(-d\tau^2+d\vtheta^2)+Q_e^2d\Omega_2^2\,,\nonumber\\
{\cal A}&=&Q_e\tan\vtheta d\tau\,,\qquad A\,=\,\sqrt{3}Q_e\tan\vtheta d\tau\,.\labell{4dim}
\eeqa
 The action \reef{act2} is invariant under global $SO(2)$ transformation under which  $({\cal A}\,, A)$ is a doublet. Using this symmetry, one can write \reef{act2} as \reef{tree} and consequently the above solution can be written as  the four dimensional compactification solution, \ie  
\beqa
ds_4^2&=& \frac{Q_e^2}{\cos^2\vtheta}(-d\tau^2+d\vtheta^2)+Q_e^2d\Omega_2^2\,,\nonumber\\
{\cal A}&=&2Q_e\tan\vtheta d\tau\,,\qquad A\,=\,0\,.
\eeqa
It is important to note that, one can not combine the whole $U(1)$ gauge field $2Q_e\tan\vtheta d\tau$ with the  metric to write a five dimensional metric. That would not satisfy the five dimensional equations of motion.

Using the coordinate transformation $\cos\vtheta=1/u$ where $1\leq u \le \infty$, the  five dimensional metric \reef{sol5} becomes
\beqa
ds_5^2&=&\rho^2\left\{-u^2d\tau^2+\frac{du^2}{u^2-1}+d\Omega_2^2\right\}+(dy+Q_e\sqrt{u^2-1}d\tau)^2\,,\labell{global4}
\eeqa
where $\rho=Q_e$. This metric  has the isometry group of $SL(2,R)\times SO(3)\times U(1)$. The Killing vector that generates  the rotational $U(1)$ isometry group is
 \beqa
\z^{(y)}&=&-\prt_y\,,
\eeqa
the Killing vectors that generate the $SO(3)$ isometry group are the following:
\beqa
{\hat{\z}}_1&=&\sin\phi\,\prt_{\theta}+\cot\theta\cos\phi\,\prt_{\phi}\,,\nonumber\\
{\hat{\z}}_2&=&-\cos\phi\,\prt_{\theta}+\cot\theta\sin\phi\,\prt_{\phi}\,,\nonumber\\
{\hat{\z}}_3&=&-\prt_{\phi}\,,
\eeqa
and the Killing vectors that generate the $SL(2,R)$ isometry group are the following:
\beqa
\z_1&=&\frac{2\sin\tau\sqrt{u^2-1}}{u}\prt_{\tau}-2\cos\tau\sqrt{u^2-1}\prt_u+\frac{2Q_e\sin\tau}{u}\prt_{y}\,,\nonumber\\
\z_2&=&\frac{2\cos\tau\sqrt{u^2-1}}{u}\prt_{\tau}+2\sin\tau\sqrt{u^2-1}\prt_u+\frac{2Q_e\cos\tau}{u}\prt_{y}\,,\nonumber\\
\z_3&=&2Q_e\prt_{\tau}\,.
\eeqa
At the boundary, $u\rightarrow \infty$,   the above Killing vectors become
\beqa
\z_{\eta}&=&\eta(\tau)\prt_\tau-\prt_\tau(\eta(\tau))u\prt_u\,,
\eeqa
for $\eta(\tau)=2\sin\tau, 2\cos\tau, 2Q_e$. If one perturbs the background \reef{global4}, then the Killing vectors will change and hence their values at the boundary will be modified. 

\section{The CFT dual }

We now study the two dimensional CFT dual of the above five dimensional solution \reef{global4} using the Brown-Henneaux technique \cite{Brown:1986nw} which makes use of the asymptotic symmetry group.
The asymptotic symmetry group  of a spacetime is the group of non-trivial allowed symmetries. A non-trivial  allowed symmetry is the one which generates a transformation that obeys the boundary conditions and its associated charge is non-vanishing \cite{Guica:2008mu}. 

Since $\prt_\tau$ is the generator whose conjugate conserved charge measures the deviation of  the solution from extremality \cite{Guica:2008mu}, we consider the perturbations that their associated conserved charges  commute with $\prt_\tau$. For the fluctuations of the metric  \reef{global4} we choose the following boundary condition:
\beqa
h_{\mu\nu}\sim{\cal O}\pmatrix{u^2&u& u& 1/u^2&1\cr 
&1& 1& 1/u&1\cr
&& 1& 1/u&1\cr
&&& 1/u^3&1/u\cr
&&&&1},\,
 \labell{M22} \eeqa 
in the basis $(\tau,\phi,\theta,u,y)$.  At the leading order, the  diffeomorphisms which preserve the above boundary condition are 
\beqa
\z_{\epsilon}&=&\epsilon(y)\prt_{y}-u\epsilon'(y)\prt_u\labell{diffeo2}\,,\\
\z^{(\tau)}&=&\prt_{\tau}\,,\nonumber\\
{\hat{\z}}_1&=&\sin\phi\,\prt_{\theta}+\cot\theta\cos\phi\,\prt_{\phi}\,,\nonumber\\
{\hat{\z}}_2&=&-\cos\phi\,\prt_{\theta}+\cot\theta\sin\phi\,\prt_{\phi}\,,\nonumber\\
{\hat{\z}}_3&=&-\prt_{\phi}\,,\nonumber
\eeqa
where $\epsilon(y)$ is an arbitrary smooth function. The Lie derivative of metric \reef{global4} with respect to $\z^{(\tau)}$ and ${\hat{\z}}$'s are zero, and  with respect to $\z_{\epsilon}$ is
\beqa
\delta_{\epsilon}ds^2&=&2\bigg((\rho^2-Q_e^2)u^2\epsilon'(y)d\tau^2+\frac{\rho^2}{(u^2-1)^2}\epsilon'(y)du^2+
\epsilon'(y)d{y}^{2}\nonumber\\
&-& \frac{Q_e}{\sqrt{u^2-1}}\epsilon'(y){d\tau} dy -\frac{\rho^2u}{u^2-1}\epsilon''(y) du dy \bigg)\,,
\eeqa
which is consistent with the boundary condition \reef{M22}. 

Using the periodicity of the $y$ coordinate, one  can expand $\epsilon(y)$ in terms of the basis $\epsilon_n(y)=-e^{-in y}$. Defining the generators $\z_n\equiv\z_{\epsilon_n}$ 
one finds they satisfy the following Virasoro algebra:
\beqa
i[\z_m,\z_n]_{L.B.}&=&(m-n)\z_{m+n}\,.
\eeqa
They have zero central charge. To evaluate the central term of the above algebra, one needs to construct the surface charges which generate the asymptotic symmetry \reef{diffeo2}. For asymptotically AdS spacetimes, the charge differences between  $(g_{\mu\nu})$ and $ (g_{\mu\nu}+h_{\mu\nu})$ are given by \cite{Barnich:2001jy} (see  \cite{Compere:2009dp} for a review)\footnote{It has been argued in \cite{Compere:2009dp} that when metric is in the canonical form the scalars and gauge fields have no contribution to the central charge in four and five dimensions. We have explicitly calculated these contribution and find zero result in the present case. }
\beqa
Q_{\z}[g]&=&\frac{1}{8\pi G}\int_{\prt\Sigma}k^{grav}_{\z}[h;g]\,,\labell{Q}
\eeqa
where the integral is over the boundary and 
\beqa
k^{grav}_{\z}[h,g]&=&-\frac{1}{2}\frac{1}{3!}\epsilon_{\alpha\beta\gamma\mu\nu}\bigg[\z^{\nu}D^{\mu}h^{\sigma}_{\sigma}-\z^{\nu}D_{\sigma}h^{\mu\sigma}+\z_{\sigma}D^{\nu}h^{\mu\sigma}+\frac{1}{2}h^{\sigma}_{\sigma}D^{\nu}\z^{\mu}\nonumber\\
&&-h^{\nu\sigma}D_{\sigma}\z^{\mu}+\frac{1}{2}h^{\sigma\nu}(D^{\mu}\z_{\sigma}+D_{\sigma}\z^{\mu})\bigg]dx^{\alpha}\wedge dx^{\beta}\wedge dx^\gamma\,,\labell{kz}
\eeqa
The covariant derivatives and raised indices are computed using the metric $g_{\mu\nu}$. 
For the diffeomorphism \reef{diffeo2}, one finds\footnote{Note that the Lie derivative of metric with respect to the diffeomorphisms $\z^{(\tau)}=\prt_{\tau}$ and ${\hat{\z}}_1, \,{\hat{\z}}_2,\,{\hat{\z}}_3$ are zero, hence, their corresponding charges are zero too.} 
\beqa
k^{grav}_{\zeta_\epsilon}&=&-
\frac{Q_e\sin\theta}{4u^2}\bigg[
2\epsilon(y)u^3\partial_y{h_{u y}} +\frac{\rho^2+Q_e^2}{\rho^2}{u}^{2}\epsilon(y){h_{y y}}\nonumber\\
&&-2\epsilon'(y){u}^{3}{h_{u y}}
+\frac{1}{\rho^2}\epsilon(y){h_{\tau\tau}}\bigg]d\theta\wedge d\phi\wedge dy\,,\nonumber
\eeqa
where we have   discarded total $\phi$ derivative terms and keep only terms that are non-zero at the boundary $u\rightarrow\infty$. We have also included only the terms that are tangent to $\prt \Sigma$.

The algebra of the non trivial asymptotic symmetries is the Poisson bracket algebra of the charges \cite{Barnich:2001jy}
\beqa
\{Q_{\z_m},Q_{\z_n}\}_{P.B.}&=&Q_{[\z_m,\z_n]}+\frac{1}{8\pi G}\int_{\prt\Sigma}k^{grav}_{\z_m}[{\cal L}_{\z_n}g,g]\,.\labell{kz3}
\eeqa
 The last term has the structure
\beqa
\frac{1}{8\pi G}\int_{\prt\Sigma}k^{grav}_{\z_m}[{\cal L}_{\z_n},g]&=&-iA(m^3+Bm)\delta_{m+n,0}\,.
\eeqa
If one defines the quantum version of the $Q$'s by
\beqa
L_n&\equiv&Q_{\z_n}+\frac{1}{2}(AB+A)\delta_{n,0}\,,
\eeqa
plus the usual rule of $\{.,.\}_{P.B.}\rightarrow -i[.,.]$, then the  algebra becomes the standard Virasoro algebra 
\beqa
[L_m,L_n]&=&(m-n)L_{m+n}+Am(m^2-1)\delta_{m+n,0}\,,\labell{viro}
\eeqa
with central charge $c=12A$. 

The  Lie derivatives of metric \reef{global4} at the boundary are
\beqa
{\mathcal{L}}_{\zeta_n} g_{\tau\tau} &=& 2i(\rho^2-Q_e^2)u^2ne^{-iny}\,,\cr 
{\mathcal{L}}_{\zeta_n} g_{\tau y}&= & -\frac{i Q_e}{u} n e^{-i n y}\,,\cr
{\mathcal{L}}_{\zeta_n} g_{u y}&= & -\frac{\rho^2}{u} n^2 e^{-i n y}\,, \cr
{\mathcal{L}}_{\zeta_n} g_{y y}& =& 2 i n e^{-i n y}\,,\cr
{\mathcal{L}}_{\zeta_n} g_{uu}& =& \frac{2 i\rho^2 }{u^4} ne^{-i n y}\,.
\eeqa
 Inserting the  above perturbation into the central term of \reef{kz3}, one finds 
\beqa
\frac{1}{8\pi G^{(5)}}\int_{\partial\Sigma}k^{grav}_{\zeta_m}[{\mathcal{L}}_{\zeta_n} g,g]=-\frac{i}{2} Q_e(m^3\rho^2+m)\delta_{m+n,0}\,\,,
\eeqa
where we have used the fact that in five dimension $G^{(5)}=2\pi$. 
Therefore, the central charge is
\beqa
c&=&6 Q_e\rho^2\,,\labell{central}\\
&=&6Q_e^3\,.\nonumber
\eeqa
This is the central charge of the CFT dual of the background \reef{global4}  in which the parameter $Q_e$ is the four dimensional electric charge. 
This central charge has been also found in \cite{Hartman:2008pb} by combining the gauge field of extremal Kerr-Newman-AdS black hole  with the 4-dimensional metric  and taking $J\rightarrow 0$. 

The Cardy formula gives the microscopic entropy of a unitary CFT at large $T_e$ to be
\beq
S_{\rm micro}=\frac{\pi^2}{3}cT_e\,.
\eeq
Using \reef{Te} and \reef{central}, one finds
\beq
S_{\rm micro}=\pi Q_e^2\,.
\eeq
This exactly reproduces the macroscopic entropy \reef{entropy2}.

\section{Discussion}
In this paper we have studied the CFT dual of the extremal  RN black hole. It has been speculated that there are two extremal limits for the RN black hole \cite{Carroll:2009ma}: The extremal RN black hole and the compactification solution. The entropy of the RN black hole is speculated in \cite{Carroll:2009ma} to be carried by the compactification solution at the extremal limit. By uplifting the compactification solution to a five dimensional solution of the Einstein-Maxwell theory, we have found the central charge of the  CFT dual of the compactification solution using the Brown-Henneaux technique, and  its microscopic entropy using the Cardy formula. The result is exactly the same as the macroscopic entropy of the RN black hole at the extremal limit. 

It has been shown in \cite{Hartman:2008pb} that the gauge fields have no direct contribution to the central charge when the metric is in the  canonical form. We have seen that only a part of the gauge field of the compactification solution (see equation \reef{4dim}) should be combined with the four dimensional metric to write the five dimensional metric in the canonical form \reef{sol5}. This indicates that only this  part of the gauge field has direct contribution to the central charge. The other part has indirect contribution  as it is  needed in order  the metric satisfies the equations of motion.

We have found the central charge of the CFT dual of the gravity on the background \reef{global4} to be given by \reef{central}. In \cite{Hartman:2008pb}, the same central charge has been found for the CFT dual of the gravity on four dimensional background of the extremal Kerr-Newman-AdS black hole fibered with a $U(1)$ gauge field at the limit of $J\rightarrow 0$. Moreover, using the Brown-Henneaux technique one can  find the  central charge for the CFT dual of the gravity on the following four dimensional metric:
\beqa
ds_4^2&=&-Q_e^2u^2d\tau^2+\left(\frac{Q_e^2}{u^2-1}\right)du^2+Q_e^2d\theta^2+Q_e^2\sin^2\theta(d\phi-Q_e\sqrt{u^2-1}d\tau)^2\label{guess}
\eeqa
The result  is exactly the same as \reef{central}. This may indicate that  the CFT dual to the gravity on these different backgrounds have the same central charge, however, other properties of the CFT theories may not be the same. It would be interesting to further study  these backgrounds to reveal   which one is corresponding to the RN black hole. The first criterion is that they must satisfy the equations of motion. Our  five dimensional solution has been found by requiring it to  satisfy the equations of motion, but the metric in (\ref{guess}) is not satisfying the equations of motion. One may also note that the  black hole solution considered in \cite{Hartman:2008pb} is not coming from a consistent Kaluza-Klein reduction, so the equations of motion are not satisfied.

{\bf Acknowledgments}: This work was  supported by  Ferdowsi University of Mashhad under grant p/18(1388/2/14). 

\end{document}